\documentclass[conference]{IEEEtran}
\IEEEoverridecommandlockouts
\usepackage{cite}
\usepackage{amsmath,amssymb,amsfonts}
\usepackage{algorithmic}
\usepackage{graphicx}
\usepackage{textcomp}
\usepackage{xcolor}
\usepackage{fancyhdr}
\usepackage{lipsum}% generate text for the example
\usepackage{booktabs}

\def\BibTeX{{\rm B\kern-.05em{\sc i\kern-.025em b}\kern-.08em
    T\kern-.1667em\lower.7ex\hbox{E}\kern-.125emX}}

\begin{document}
\title{Improved Leakage Abuse Attacks in Searchable Symmetric Encryption with eBPF Monitoring}

\author{\IEEEauthorblockN{Chinecherem Dimobi}
\IEEEauthorblockA{\textit{Virginia Tech} \\
chinedimobi@vt.edu}
}

\maketitle

\begin{abstract}
Searchable Symmetric Encryption (SSE) allows users to search over encrypted data stored on untrusted servers, like cloud providers. While SSE hides the content of queries and documents, it still leaks patterns, such as how often a query is made. These leakages have been shown to enable leakage abuse attacks, but recent defenses have made such attacks harder to carry out.
In this work, we explore how system-level monitoring using eBPF (Extended Berkeley Packet Filter) can be used to uncover new forms of leakage that go beyond what is typically captured in SSE threat models. By observing low-level system behavior during search operations, we show that an attacker can gain additional insights into query behavior, document access, and processing flow. We define a new leakage pattern based on these observations and demonstrate how they can strengthen existing attacks.
Our findings suggest that system-level leakages present a practical threat to SSE deployments and must be considered when designing defenses. This work serves as a step toward bridging the gap between theoretical SSE security and the realities of system-level exposure.
\end{abstract}


\section{Introduction}
Searchable Symmetric Encryption (SSE) makes it possible to search over encrypted data stored on external servers \cite{curtmola2006searchable}. This has become an important tool for preserving privacy in outsourced storage, especially when using cloud services. The basic idea is that a user can upload encrypted files and still perform keyword searches without revealing the content of the files or the queries. However, the security guarantees of SSE are still limited. SSE systems are known to leak certain types of information, like how many times a query is made or how many documents it matches. These leakages can be exploited through what are called leakage-abuse attacks.

One common leakage-abuse attack is the Frequency Matching Attack (FMA) \cite{oya2021hiding}, where an attacker observes the frequency of search queries and correlates them with known keyword distributions to infer the search terms. But research has made progress in defending against such attacks \cite{xu2021interpreting}, with strategies like padding query results or hiding volumes. These improvements have made it harder for an attacker to learn much just from query patterns.

This project takes a step back and asks a different question: What happens if an attacker has access not just to the query patterns, but to the low-level behavior of the system itself? Specifically, what if the attacker can monitor what the kernel is doing during a search operation? That is where eBPF (Extended Berkeley Packet Filter) comes in. eBPF \cite{jia2023programmable, gbadamosi2024ebpf} is a Linux kernel extension that allows programs to run inside the kernel and observe or respond to system events in real time. It is often used for debugging and performance monitoring, but it can also be used to track things like which files are opened and in what order.

The goal of this research is to explore whether system-level monitoring via eBPF can reveal new leakages in SSE systems that existing defenses do not account for. In doing so, we aim to:
\begin{itemize}
  \item Investigate what low-level events occur during SSE search operations.
  \item Identify any additional leakage patterns that can be observed from these events.
  \item Show how this information can be used to enhance or create more effective leakage abuse attacks.
\end{itemize}

Our contributions in this work are as follows:
\begin{itemize}
  \item We identify and define a new leakage pattern, \(L_{fileAccess}\), that reveals the exact files accessed during a query.
  \item We demonstrate how this leakage can be used to recover queries more accurately, even when traditional frequency-based attacks fail.
  \item We highlight the gap between theoretical SSE security models and real-world deployment vulnerabilities, suggesting directions for future defenses.
\end{itemize}

This work aims to encourage a broader view of SSE security to include what can be observed at the system level during search operations at runtime.

\section{Related Works}

\subsection{Evolution of Searchable Symmetric Encryption (SSE) Schemes}

Searchable Symmetric Encryption (SSE) was first introduced by Song et al. \cite{song2000practical}, who proposed a method for performing keyword searches directly over encrypted data. Their work proved it was possible to preserve privacy while enabling some form of search. However, their scheme had limitations: it required scanning every file linearly and did not clearly define what information might be unintentionally leaked.

Curtmola et al. \cite{curtmola2006searchable} later proposed a more structured approach to SSE. They introduced clear security definitions and formalized how to reason about leakage. Their scheme separated the encrypted index from the encrypted data, making searches faster and easier to manage. This work became the baseline for many future designs.

Storage systems moved to the cloud and real-world datasets became more dynamic which led to Dynamic SSE (DSSE), which supports updates such as adding or removing documents. Kamara et al. \cite{kamara2012dynamic} proposed one of the first DSSE schemes. Their work allowed encrypted data to be updated while preserving security, but also introduced new leakage vectors.

Subsequent works focused on making SSE schemes more expressive and scalable, supporting richer queries, multi-client environments, and performance at scale \cite{cash2014dynamic}. But as SSE systems grew in capability, the complexity of managing and understanding leakage also increased leading to the trade-off between functionality and privacy.

\subsection{Leakage in SSE Systems}
Leakages in search operations mainly come as access patterns, which tell which documents are returned for a given query, and search patterns, which show whether a particular query has been repeated.

Islam et al. \cite{islam2012access} showed that observing access patterns over time combined with background knowledge about the data, could recover the original queries. Oya and Kerschbaum \cite{oya2021hiding} added that hiding access patterns is not enough, as repeated queries can still leak information. More recently, Xu et al. \cite{xu2023leakage} showed that even forward- and backward-private systems leak data through volume and co-occurrence. These subtle leakages, like how often documents appear together, can still help attackers reconstruct query intent.

\subsection{Leakage-Abuse Attacks}
Leakage-abuse attacks exploit these leakage patterns without breaking the underlying encryption. Cash et al. \cite{cash2015leakage} introduced attacks using frequency analysis to map encrypted queries to known keyword distributions.

Gui et al. \cite{gui2023volume} explored volume leakage, where attackers use the number of returned documents to infer query intent. Bakas and Michalas \cite{bakas2023forward} demonstrated attacks in multi-client SSE systems, where shared index structures introduce cross-client leakage.

These studies show that cryptographic protections are not enough when attackers can exploit what the system leaks through normal operation. This motivates broader threat models that account for how SSE interacts with its environment.

\subsection{System-Level Monitoring with eBPF}
eBPF (Extended Berkeley Packet Filter) \cite{ebpf_overview} is a Linux kernel technology originally built for packet filtering. It has grown into a powerful framework for observing and reacting to system events \cite{gregg2019bpf, borkmann2015contain}. It allows small programs to run safely in the kernel, enabling low-overhead tracing, syscall filtering \cite{jia2023programmable}, container security \cite{bpfcontain2021}, and network monitoring \cite{cilium2020cilium}.

Tools like Cilium and Tracee use eBPF to provide deep runtime visibility. Because eBPF operates within the kernel with strict safety checks, it can capture detailed system behaviors without compromising performance \cite{ebpf2023observability, tracee2022}.

Recent work by Lu et al. \cite{lu2024moat} and Nguyen et al. \cite{nguyen2023systematic} shows that system-level patterns, like file access and I/O operations, introduce leakage not captured by cryptographic threat models.

Our research builds on this idea. We explore whether attackers could use system-level visibility, through tools like eBPF, to infer sensitive patterns during SSE operations. For example, observing file access calls could reveal how encrypted queries interact with data, creating a new class of leakages not considered in existing models. Our findings push toward a more holistic view of SSE security that accounts for what the system itself reveals during encrypted search.

\section{Preliminaries}

\subsection{How SSE Works: A Technical Overview}

Symmetric Searchable Encryption (SSE) schemes are designed to enable keyword searches over encrypted documents stored on untrusted platforms such as cloud servers. The key idea is that the cloud should be able to respond to search queries without learning the contents of the documents, the keywords being searched, or the results being returned.

\subsubsection{Inverted Index Construction}

The process begins on the data owner’s side. All documents are parsed to extract keywords. These keywords are then used to build an inverted index: a data structure that maps each keyword to the list of document identifiers (e.g., file IDs) where the keyword appears. For example:

\begin{verbatim}
"invoice"  →  {doc1, doc3, doc7}
"contract" →  {doc2, doc4}
\end{verbatim}

This index is crucial because it supports efficient retrieval, rather than scanning all encrypted files for every search.

\subsubsection{Index and Document Encryption}

Each document is encrypted using a symmetric encryption scheme. Simultaneously, the inverted index is encrypted using deterministic or structured encryption so that search tokens can match index entries without revealing plaintext content. The encrypted index and encrypted documents are then uploaded to the Cloud Service Provider (CSP). The CSP stores these without having the ability to decrypt them.

\subsubsection{Search Protocol}

To perform a search, the client (data owner or authorized user) uses a secret key to generate a deterministic search token corresponding to the keyword of interest. This token, usually denoted $t_w$ for a keyword $w$, is sent to the CSP.

The CSP uses $t_w$ to search the encrypted index and find the encrypted document identifiers linked to that keyword. Once found, the CSP returns the encrypted files corresponding to those identifiers.

The client then decrypts the documents locally using their secret key. At no point is the keyword or the decrypted document revealed to the CSP.

\subsubsection{Update Protocols (Add/Delete)}

Dynamic SSE (DSSE) supports updates to the encrypted database, including adding and deleting documents.

\begin{itemize}
  \item \textbf{Add:} When a new document is added, the client parses the document to extract keywords, updates the local inverted index, encrypts both the document and the new index entries, and sends them to the CSP.
  \item \textbf{Delete:} When a document is deleted, the client sends a deletion token to the CSP, which removes the corresponding encrypted document and updates the affected index entries.
\end{itemize}

To maintain security, update operations also require cryptographic protocols that prevent the CSP from learning what has changed, aside from the number of entries affected.

\subsubsection{Leakage Types}

Each SSE operation leaks a slightly different kind of metadata:

\begin{itemize}
  \item \textbf{Add ($L_{add}$)} leaks the number of keywords in the new document being added.
  \item \textbf{Update ($L_{update}$)} leaks the number of index entries modified or added.
  \item \textbf{Delete ($L_{delete}$)} leaks the number of index entries being removed.
  \item \textbf{Search ($L_{search}$)} leaks the number of documents returned for a given keyword.
\end{itemize}

Most of this leakage stems from the fact that encrypted identifiers in the index remain linkable over time, even if their contents are hidden.

\subsection{Frequency Matching Attack (FMA)}

The Frequency Matching Attack (FMA) is a classic leakage-abuse technique that takes advantage of the repeated structure in queries and responses. The attacker begins by observing a sequence of search tokens and recording how frequently each token is used (i.e., query frequency) and how many documents each token retrieves (i.e., result size). This creates a frequency distribution of encrypted queries.

If the attacker has access to an auxiliary dataset (e.g., a similar plaintext dataset or prior knowledge), they can compute the frequency distribution of known keywords in that dataset. By matching the two distributions, the attacker tries to recover the mapping between encrypted search tokens and plaintext keywords.

The success of FMA improves significantly when the attacker can observe the result sizes ($L_{search}$) of the searches. Keywords with unique result sizes can often be matched one-to-one with specific search tokens, even if their frequency is ambiguous.

However, FMA is not perfect. One of its main limitations is when multiple keywords share similar frequency distributions. For example, if “invoice,” “contract,” and “budget” each appear in ~20\% of the documents, then their tokens will have similar frequency and result size profiles. In this case, FMA cannot confidently link tokens to keywords as there is no unique mapping.

This limitation leaves open the possibility of new vectors. If frequency is not enough, could something else, like system-level information reveal what FMA cannot? That question is the starting point for our work.

\subsection{How eBPF Works}
The kernel is a core component of any system that manages all the resources for that system. The Extended Berkeley Packet Filter (eBPF) is a Linux kernel technology that allows running small, sandboxed programs directly inside the kernel without modifying its source code or loading custom kernel modules. eBPF programs can interact safely with various kernel components in real time, providing visibility into system behaviors and performance.

\subsubsection{Core Concepts of eBPF}

eBPF operates by attaching small programs to specific kernel events. These events might include system calls (like \texttt{open}, \texttt{read}, or \texttt{write}), network events (such as incoming packets), or even custom tracepoints defined within the kernel.

Each eBPF program is verified by a built-in safety checker, known as the \textit{eBPF verifier}, before it executes. This verifier ensures that the eBPF program will nnot crash the kernel, completes execution in a bounded amount of instructions and only accesses authorized kernel memory. Once verified, the program runs when triggered by a kernel event. When an event occurs, the eBPF program executes and may log or analyze relevant information.

\section{Threat Model and Security Assumptions}

% \subsection{Entities and Roles}

% Our system involves three main entities:

% \begin{itemize}
%     \item \textbf{Data Owner (Client)}: Owns and encrypts the dataset, builds the SSE index, and issues search queries.
%     \item \textbf{Cloud Service Provider (CSP)}: Stores encrypted data and the encrypted index. The CSP processes search and update requests on behalf of the client.
%     \item \textbf{Adversary}: An external entity whose goal is to infer sensitive information from observed operations.
% \end{itemize}

\subsection{Adversary Model}

We assume an \textbf{honest-but-curious} adversary model. In other words, the CSP faithfully executes all protocols and does not maliciously alter the stored data or queries. However, the CSP is curious as it tries to learn as much as possible from the information naturally revealed through processing queries and managing data.

Specifically, the adversary has the following capabilities:

\begin{enumerate}
    \item \textbf{Full visibility into encrypted operations:} The attacker sees all encrypted documents, encrypted indexes, and query tokens processed by the CSP.
    
    \item \textbf{System-level monitoring access:} Using tools like eBPF, the attacker can observe the detailed behavior of the CSP’s system, such as which encrypted files are accessed during each operation, the sequence of system calls, and possibly even timing information.
    
    \item \textbf{Auxiliary (background) knowledge:} The attacker has partial knowledge about the underlying dataset, such as keyword frequency distributions or statistics about the data. This knowledge could come from a similar plaintext dataset or publicly available sources. The adversary has partial knowledge about the plaintext index. This includes knowing some document distributions.
\end{enumerate}

However, the adversary cannot directly break the underlying cryptographic primitives. The encryption schemes used are secure, and the secret keys remain confidential to the data owner.

\subsection{Security Assumptions}

To clearly frame the attacker’s perspective, we rely on the following assumptions:

\begin{itemize}
    \item All keywords, documents, and indexes are encrypted using secure, well-established cryptographic methods. The adversary cannot directly decrypt or invert encryption schemes without the secret keys.
    
    \item Document identifiers within the encrypted index are treated as opaque. In traditional SSE models, these IDs should not reveal meaningful information about the documents they represent.
    
    \item The communication between the data owner and the CSP is protected. The attacker sees only the resulting encrypted queries and responses, not the plaintext communication channel.
\end{itemize}

\subsection{Attack Goal}

The primary goal of our adversary is to recover plaintext search queries issued by the client.

We focus particularly on leakage-abuse attacks, like the Frequency Matching Attack (FMA), that leverage metadata leaks, such as  how many documents are returned per query.

\subsection{Scope and Limitations}

In our threat model, we do not consider an attacker who can modify queries or stored data actively. Our model focuses solely on passive attacks, aligning with realistic cloud scenarios where the CSP executes protocols correctly but tries to infer additional information passively.

% \section{Methodology}
% \subsection{Rationale for System-Level Leakage}

% Traditional SSE threat models typically assume leakages at the cryptographic or application layer only. However, real systems may inadvertently leak extra information at the system-level, such as file-access patterns or system-call sequences, that traditional security models don't account for.

% Tools like eBPF make these previously hidden system behaviors visible. The attacker, using eBPF, can monitor which specific encrypted files are accessed each time a search query occurs, potentially revealing additional leakage vectors.

\section{Methodology}

In this section, we present the approach we took to investigate whether additional leakages from system-level monitoring can significantly strengthen leakage-abuse attacks in SSE, particularly focusing on how eBPF monitoring can help attackers infer queries more accurately.

\subsection{Intuition and High-Level Idea}

Our intuition builds on the fact that standard SSE schemes often overlook leakages that appear when the CSP interacts with encrypted data at the system level. For instance, when the CSP processes a search query, it has to access certain encrypted files from storage and return them to the client. Even though these files and indexes are encrypted, the filenames themselves are usually not changed. This means filenames visible at the operating system level correspond directly to the original plaintext filenames.

We leverage eBPF to observe exactly which encrypted files are accessed during each search. Because eBPF can safely monitor kernel events, we can easily track the filenames of ciphertext files accessed during a query. Since we assume the attacker has partial knowledge of the plaintext dataset and index, observing these filenames directly provides a new leakage vector that was not thoroughly considered before. This approach allows the attacker to match the filenames from encrypted queries to known plaintext filenames, revealing the relationship between encrypted tokens and plaintext keywords.

\subsection{Defined Leakage Vector \( L_{fileAccess} \)}

Our approach introduces a new system-level leakage pattern called \( L_{fileAccess} \). The main idea behind \( L_{fileAccess} \) is simple yet powerful: 

\begin{enumerate}
    \item Each time a client sends a search query to the CSP, the CSP processes the query by looking up the encrypted index and then accessing specific ciphertext files. 
    \item By attaching eBPF monitoring programs directly into the kernel at critical system calls (e.g., \texttt{openat}, \texttt{read}), we can log exactly which ciphertext files are accessed for each query.
\end{enumerate}

Formally, we define our new leakage vector \( L_{fileAccess} \) as follows:
    
    For each search token $t_w$, the leakage reveals the set:
    \[
    t_{w} \rightarrow \{f_{1}, f_{2}, \dots, f_{n}\}
    \]
    Here, \( f_{i} \) represents the filenames of ciphertext documents that are accessed and returned to the client as part of the query $t_w$.

This new leakage vector complements traditional leakage such as result size and frequency patterns. Instead of just knowing how many files a query returns, the attacker now knows exactly which ciphertext files correspond to each search.

\subsection{Enhanced Frequency Matching Attack (eFMA)}

To practically leverage the new leakage vector (\( L_{fileAccess} \)), we extend the standard Frequency Matching Attack (FMA) to what we call eBPF Enhanced Frequency Matching Attack (eFMA). The goal of eFMA is straightforward, to recover the plaintext keyword behind each encrypted search query. The procedure for eFMA goes as follows:

\begin{itemize}
    \item First, observe each query token and log two critical pieces of information:
    \begin{enumerate}
        \item The \textit{result length} (the number of files returned) for each query.
        \item The exact set of encrypted documents accessed for that query, captured using eBPF monitoring.
    \end{enumerate}
    \item Next, build a candidate mapping for each query token. This is done by comparing the observed ciphertext files from each query (\( F_{e} \)) to the plaintext filenames known from partial knowledge (\( F_{p} \)).
    \item We perform frequency analysis first. Queries with unique frequencies (distinct result lengths) can be directly matched to plaintext keywords. However, when multiple keywords have similar frequencies, traditional FMA struggles to differentiate them.
    \item When frequencies tie, we leverage the exact sets of ciphertext filenames accessed during queries. If the ciphertext filenames match exactly with filenames in the plaintext index (\(F_{e} = F_{p}\)), we directly infer the token’s keyword.
    \item Formally, this mapping can be expressed as:
    \[
    t_{w} = w \quad \text{if} \quad F_{e} = F_{p}
    \]
    where $t_{w}$ is the encrypted token for keyword \( w \), and \( F_{e} \), \( F_{p} \) are the sets of ciphertext and plaintext files respectively.
\end{itemize}

This enhanced matching approach greatly improves our capability to recover queries compared to traditional frequency-based attacks alone.

\subsection{Protocol Analysis}

Our protocol leverages practical observations made possible by eBPF to significantly strengthen traditional leakage abuse attacks. A critical insight here is that SSE schemes generally do not rename or obfuscate filenames at the file system level. This small oversight creates a powerful leakage channel.

From an adversarial standpoint, this means even forward-private or theoretically secure SSE schemes remain vulnerable to system-level leakages like file access patterns. It becomes clear that security models need to account explicitly for such leakages, beyond just cryptographic guarantees or standard index-level leakages.

In summary, by integrating system-level file access monitoring through eBPF, our enhanced approach significantly advances leakage abuse attacks, highlighting an important and often overlooked dimension of SSE security.

\section{Implementation and Experiments}

\subsection{System Setup and Configuration}

Our implementation was carried out using the DK-Nguyen DSSE scheme \cite{nguyen_searchable_encryption} with forward privacy, which follows a standard three-party model involving a Data Owner, a Trusted Authority, and a Cloud Service Provider (CSP). The data owner encrypts a set of documents and builds the corresponding encrypted index. The encrypted data is then uploaded to the CSP, where queries are later executed.

We ran the DSSE scheme in a Docker container on a virtual machine and used a subset of 100 emails from the Enron \cite{enron_dataset} dataset as the experimental corpus. This dataset was chosen for its real-world relevance and manageable size, making it suitable for observing query-level leakage. We limited our evaluation to search-only queries (no updates) in order to focus purely on the leakage induced during retrieval.

To trace system-level activity during these queries, we used \texttt{bpftrace} \cite{bpftrace}, a dynamic tracing tool built on top of eBPF. \texttt{bpftrace} allowed us to monitor file-level system calls, particularly those triggered during document access, without modifying the DSSE codebase.

\subsection{Evaluation Metrics}

The primary metric used in our evaluation is \textbf{query recovery accuracy}, which measures how accurately the attacker can map encrypted search tokens to their corresponding plaintext keywords. This metric is appropriate for our work, since the core goal of both the baseline and enhanced attacks is to infer plaintext queries from observable leakage.

We compared two attacks:

\begin{itemize}
    \item \textbf{Baseline Frequency Matching Attack (FMA):} Uses only the result size (number of documents returned) for each query to perform frequency-based inference.
    \item \textbf{Enhanced Frequency Matching Attack (eFMA):} Uses both the result size and the set of ciphertext files accessed during each query, obtained through eBPF monitoring.
\end{itemize}

\subsection{Experimental Design}

To simulate the adversary, we followed a passive attack setup. We acted as the CSP, observing query tokens and the files returned for each query, without modifying or interfering with the DSSE code.

\paragraph{Keyword Selection} We queried a curated set of keywords from the Enron dataset, ensuring a mixture of frequent and infrequent terms. This created a range of frequency distributions that exposed both the strengths and weaknesses of frequency-only attacks.

\paragraph{Leakage Collection} 

\begin{itemize}
    \item For the baseline attack, we collected the result size for each query.
    \item For the enhanced attack, we used \texttt{bpftrace} to log the specific ciphertext filenames accessed during each query. These filenames were not obfuscated in the storage system, allowing us to observe exact document access patterns.
\end{itemize}

\paragraph{Attacker Knowledge} We assumed the attacker has partial access to the plaintext dataset (e.g., from a public copy of Enron) and knows the mapping between keywords and the documents they appear in. This knowledge enables them to build a candidate inverted index, which is used for frequency matching and document set comparison.

\subsection{Results and Analysis}

\begin{figure}[h!]
\centering
\includegraphics[width=0.65\linewidth]{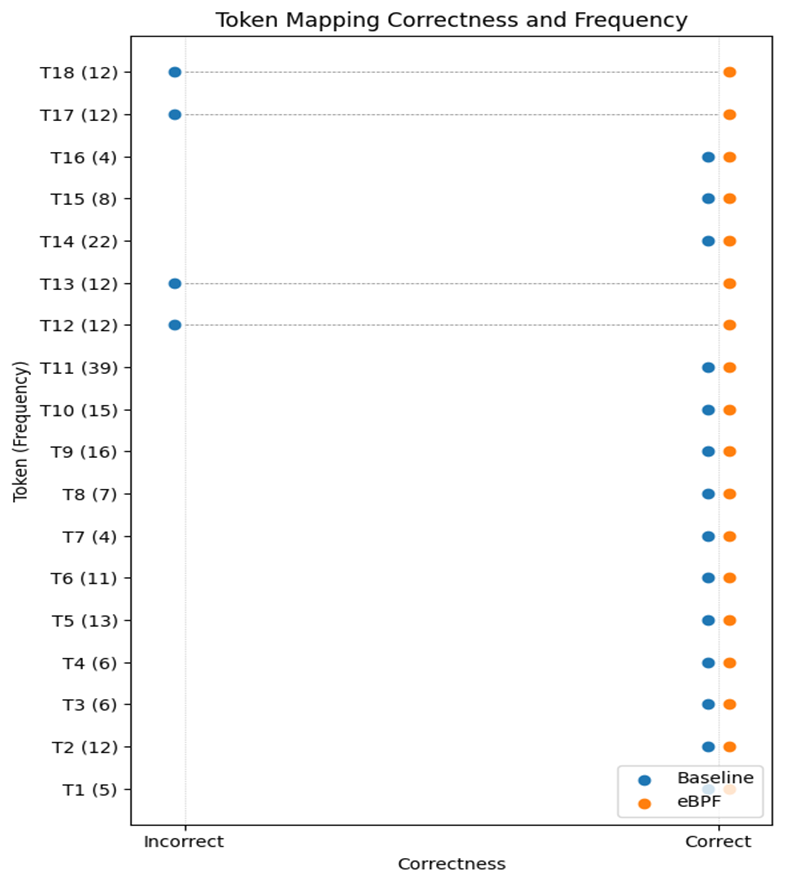}
\caption{Token mapping accuracy across queries with and without eBPF leakage}
\label{fig:accuracy}
\end{figure}

The results of our experiments are summarized in Table~\ref{tab:accuracy} and Figure~\ref{fig:accuracy}.

\begin{table}[htbp]
\caption{Keyword Recovery Accuracy Comparison}
\label{tab:accuracy}
\begin{center}
\begin{tabular}{lc}
\toprule
\textbf{Method} & \textbf{Query Recovery Accuracy} \\
\midrule
Baseline (FMA) & 77.8\% \\
Enhanced (eFMA) & 100\% \\
\bottomrule
\end{tabular}
\end{center}
\end{table}

The baseline FMA achieved a query recovery accuracy of 77.8\%. After adding system-level leakage through eBPF tracing, the enhanced attack reached 100\% accuracy.
The results of our experiment clearly show how file-level leakage captured with eBPF improves the accuracy of query recovery. Figure~\ref{fig:accuracy} shows a breakdown of each token’s mapping outcome under both the baseline Frequency Matching Attack (FMA) and our enhanced eFMA method.

Each dot represents a token. Blue dots show the baseline results, while orange dots show the results after adding eBPF-based file access leakage. 
We see that tokens like \textbf{T12}, \textbf{T13}, \textbf{T17}, and \textbf{T18} were all misclassified under the baseline attack. These tokens shared the same frequency of 12, making them indistinguishable from one another using only result sizes.

In contrast, the enhanced eFMA approach correctly mapped all tokens, including those that failed under the baseline. Even though \textbf{T12} and \textbf{T13} both had the same frequency, they accessed different sets of ciphertext files during their respective queries. Using eBPF, we were able to observe these file accesses and match them to known document sets in the plaintext index. This extra leakage broke the tie that the baseline method could not resolve.

Other tokens like \textbf{T11}, \textbf{T9}, and \textbf{T1} had unique frequencies and were correctly identified by both methods, which is expected. 

\subsection{Key Insights}

\begin{itemize}
    \item \textbf{File access patterns leak more than result sizes.} With only a few lines of eBPF tracing code, we could recover not just how many documents were returned, but exactly which ones.
    \item \textbf{System-level leakage significantly increases attack power.} The transition from 77.8\% to 100\% recovery demonstrates that file-level observation fills the gap left by frequency-based ambiguity.
    \item \textbf{The DSSE system is vulnerable without even being modified.} Our enhanced attack required no changes to the SSE scheme. We treated the entire system as a black box and extracted all leakage externally, suggesting real-world deployments could be at risk even when using formally secure SSE constructions.
\end{itemize}

\section{Future Work}

There are several interesting directions to build on this work. First, while our current attack assumes the adversary has partial access to the plaintext dataset (to help build the index), it would be useful to reduce that assumption and explore how far an attacker can go with more limited background knowledge.

Another natural next step is to run this same attack in the presence of updates. DSSE schemes often allow document insertions and deletions, and it would be valuable to see how those operations affect the leakage we can observe with eBPF. If new patterns emerge during updates, they could be just as useful for attackers.

We are also interested in seeing whether mechanisms like Oblivious RAM (ORAM) can help defend against this kind of leakage. ORAM hides access patterns by reshuffling data and adding dummy accesses, which might mitigate the signals eBPF relies on. It is an open question whether that would be enough, especially when the attacker is watching at the syscall level.

Beyond file access, eBPF gives us access to a wider range of system events. Future work can look into timing side channels, memory usage patterns, or syscall sequences to uncover even more leakages. These could be layered on top of existing attacks or used to enhance others in the leakage-abuse category.

\section{Conclusion}

This work demonstrates that system-level behavior can create powerful leakage channels in Searchable Symmetric Encryption schemes. Specifically, we show that using eBPF to observe file accesses on the CSP side gives an attacker significantly more power to recover queries.

Our enhanced attack, which uses these file-level leakages, outperforms traditional frequency-based approaches and achieves full query recovery in our test dataset. The results highlight the gap between formal security models and how real-world systems behave under the hood.

To move forward, it is important for future SSE and DSSE schemes to consider system-level leakage in their security models. Tools like eBPF make it easy for attackers to observe what the kernel sees, so these vectors can no longer be ignored. We recommend that future work includes stronger defenses against file access leakage, possibly using obfuscation techniques like ORAM, and that researchers revisit their threat models to reflect what is practically observable in deployed environments.

\bibliographystyle{IEEEtran} 
\bibliography{Mybib}





% \end{thebibliography}

\end{document}